\PassOptionsToPackage{names,dvipsnames}{xcolor} %
\documentclass[sigplan,10pt]{acmart}

\usepackage[english]{babel}
\usepackage[inference]{semantic}
\usepackage{amsmath}\allowdisplaybreaks
\usepackage{mathpartir}
\usepackage{amsthm}
\usepackage{amssymb}
\usepackage{stmaryrd} %
\usepackage{xspace}
\usepackage{latexsym}
\usepackage{hyperref}
\usepackage{ifthen}
\usepackage{mathtools}
\usepackage{color}
\usepackage{listings}
\usepackage{verbatim}
\usepackage[colorinlistoftodos]{todonotes}
\usepackage{tikz}
\usetikzlibrary{positioning,shadows,arrows,calc,backgrounds,fit,shapes,snakes,shapes.multipart,decorations.pathreplacing,shapes.misc,patterns}
\usepackage{xspace}
\usepackage[T1]{fontenc}
\usepackage[scaled=.83]{beramono}
\usepackage{epigraph}
\usepackage{cleveref}
\usepackage{booktabs}
\usepackage{float}
\usepackage{etoolbox}
\usepackage{wrapfig}
\usepackage{pgfplots}
\usepackage{nameref}
\usepackage{scalerel}
\usepackage{bm}
\usepackage{centernot}
\usepackage{mdframed}
\usepackage{natbib}
\setcitestyle{number}

\hypersetup{ pdfpagemode=UseOutlines, colorlinks=true, linkcolor=ForestGreen, citecolor=ForestGreen }

\newcommand{\mi}[1]{\ensuremath{\mathit{#1}}}

\newcommand*{\QEDA}{\hfill\ensuremath{\blacksquare}}%

\AtEndEnvironment{problem}{\null\hfill\QEDA}
\AtEndEnvironment{example}{}

\Crefname{lstlisting}{Listing}{Listings}
\Crefname{problem}{Problem}{Problems}

\Crefname{equation}{Rule}{Rules}

\newcounter{typerule}
\crefname{typerule}{rule}{rules}

\pgfdeclarelayer{background}
\pgfdeclarelayer{veryback}
\pgfdeclarelayer{veryback2}
\pgfdeclarelayer{veryback3}
\pgfdeclarelayer{back2}
\pgfdeclarelayer{foreground}
\pgfsetlayers{veryback3,veryback2,veryback,background,back2,main,foreground}

\newcounter{line}

\definecolor{mygreen}{rgb}{0,0.6,0}
\definecolor{mygray}{rgb}{0.5,0.5,0.5}
\definecolor{mymauve}{rgb}{0.58,0,0.82}

\lstdefinelanguage{Java} 
{morekeywords={abstract, all, and, as, assert, but, disj, else, exactly, extends, fact, for, fun, iden, if, iff, implies, in, Int, int, let, lone, module, no, none, not, one, open, or, part, pred, run, seq, set, sig, some, sum, then, univ, package, class, public, private, null, return, new, interface, extern, object, implements, System, static, super, try , catch, throw, throws, Unit, var, val, of, principal, trust},
sensitive=true,
keywordstyle=\bfseries\color{green!40!black},
commentstyle=\itshape\color{purple!40!black},
morecomment=[l][\small\itshape\color{purple!40!black}]{//},
identifierstyle=\color{blue},
stringstyle=\color{orange},
basicstyle=\small,
basicstyle={\small\ttfamily},
numbers=left,
numberstyle=\tiny\color{mygray},
tabsize=2,
numbersep=3pt,
breaklines=true,
lineskip=-2pt,
stepnumber=1,
captionpos=b,
breaklines=true,
breakatwhitespace=false,
showspaces=false,
showtabs=false,
float=!h,
columns=fullflexible,escapeinside={(*@}{@*)},
moredelim=**[is][\color{red!60}]{@}{@},
literate={->}{{$\to$}}1 {^}{{$\mspace{-3mu}\widehat{\quad}\mspace{-3mu}$}}1
{<}{$<$ }2 {>}{$>$ }2 {>=}{$\geq$ }2 {=<}{$\leq$ }2
{<:}{{$<\mspace{-3mu}:$}}2 {:>}{{$:\mspace{-3mu}>$}}2
{=>}{{$\Rightarrow$ }}2 {+}{$+$ }2 {++}{{$+\mspace{-8mu}+$ }}2
{<=>}{{$\Leftrightarrow$ }}2 {+}{$+$ }2 {++}{{$+\mspace{-8mu}+$ }}2
{\~}{{$\mspace{-3mu}\widetilde{\quad}\mspace{-3mu}$}}1
{!=}{$\neq$ }2 {*}{${}^{\ast}$}1 
{\#}{$\#$}1
}

\lstdefinelanguage{Asm}
{morekeywords={abstract, all, and, as, assert, but, check, disj, else, exactly, extends, fact, for, fun, iden, if, iff, implies, in, Int, int, let, lone, module, no, none, not, one, open, or, part, pred, run, seq, set, sig, some, sum, then, univ, package, class, public, private, null, return, new, interface, extern, object, implements, System, static, super, try , catch, throw, throws, Unit, var, val, principal, trust, label, load, add, addi, into, test},
sensitive=true,
identifierstyle=\color{black},
keywordstyle=\bfseries,
commentstyle=\itshape\color{purple!40!black},
morecomment=[l][\small\itshape\color{purple!40!black}]{//},
stringstyle=\color{orange},
basicstyle=\small,
basicstyle={\small},
numbers=left,
numberstyle=\tiny\color{mygray},
tabsize=2,
numbersep=3pt,
breaklines=true,
lineskip=-2pt,
stepnumber=1,
captionpos=b,
breaklines=true,
breakatwhitespace=false,
showspaces=false,
showtabs=false,
float=!h,
columns=fullflexible,escapeinside={(*@}{@*)},
moredelim=**[is][\color{red!60}]{@}{@},
literate={->}{{$\to$}}1 {^}{{$\mspace{-3mu}\widehat{\quad}\mspace{-3mu}$}}1
{<}{$<$ }2 {>}{$>$ }2 {>=}{$\geq$ }2 {=<}{$\leq$ }2
{<:}{{$<\mspace{-3mu}:$}}2 {:>}{{$:\mspace{-3mu}>$}}2
{=>}{{$\Rightarrow$ }}2 {+}{$+$ }2 {++}{{$+\mspace{-8mu}+$ }}2
{<=>}{{$\Leftrightarrow$ }}2 {+}{$+$ }2 {++}{{$+\mspace{-8mu}+$ }}2
{\~}{{$\mspace{-3mu}\widetilde{\quad}\mspace{-3mu}$}}1
{!=}{$\neq$ }2 {*}{${}^{\ast}$}1 
{\#}{$\#$}1
}
\DeclareMathOperator\ceq{\ensuremath{\mathrel{\simeq_{\mi{ctx}}}}}

\def\teqaux#1{\vcenter{\hbox{\ooalign{\hfil
       \raise6pt \hbox{\scriptsize{T}}\hfil\cr\hfil
       $=$}}}}

\def\ceqwaux#1{\vcenter{\hbox{\ooalign{\hfil
       \raise6pt \hbox{\scriptsize{w-b}}\hfil\cr\hfil
       $\ceq$}}}}

\def\praux#1{\vcenter{\hbox{\ooalign{\hfil
       \raise4pt \hbox{$\subset$}\hfil\cr\hfil
       $\sim$}}}}

\theoremstyle{definition}

\Crefname{corollary}{Corollary}{Corollaries}
\Crefname{informal}{Definition}{Definition}
\Crefname{assumption}{Assumption}{Assumptions}
\crefname{assumption}{Assumption}{Assumptions}
\Crefname{property}{Property}{Properties}
\crefname{property}{Property}{Properties}

\Crefname{paragraph}{Section}{Sections}

\newcommand{\wasm}[0]{\text{Wasm}\xspace}
\newcommand{\mswasm}[0]{\text{MS-Wasm}\xspace}

\AtEndEnvironment{example}{\null\hfill$\boxdot$}

\makeatletter
\xdef\@thefnmark{\@empty}

\makeatother

\newcounter{hps}
\crefname{hps}{}{}

\newcommand{\proven}[1]{\ensuremath{\checkmark}}

\title{
	Memory Safety Preservation for WebAssembly
}

\author{
	Marco Vassena
}
\affiliation{
	CISPA Helmholz Center for Information Security
}
\author{
	Marco Patrignani
}
\affiliation{
	Stanford University
}
\affiliation{
	CISPA Helmholz Center for Information Security
}
\date{}

\begin{abstract}
WebAssembly (\wasm) is a next-generation portable compilation target
 for deploying applications written in high-level languages on the
 web.
In order to protect their memory from untrusted code, web browser
engines confine the execution of compiled \wasm programs in
a \emph{memory-safe} sandbox.
Unfortunately, classic memory-safety vulnerabilities (e.g., buffer
overflows and use-after-free) can still corrupt the
memory \emph{within} the sandbox and allow \wasm code to mount severe
attacks.
To prevent these attacks, we study a class of secure compilers that
eliminate (different kinds of) of memory safety violations.
Following a rigorous approach, we discuss memory safety in terms of
hypersafety properties, which let us identify suitable secure
compilation critera for memory-safety-preserving compilers.
We conjecture that, barring some restrictions at module boundaries,
  the existing security mechanisms of \wasm may suffice to enforce
  memory-safety preservation, in the short term.
In the long term, we observe that certain features proposed in the
design of a memory-safe variant of \wasm could allow compilers to lift
these restrictions and enforce relaxed forms of memory safety.
\end{abstract}

\begin{document}
\maketitle

\section{Introduction}
WebAssembly (\wasm) has gained traction as the new portable compilation target
language for deploying on the web applications written in high-level languages
like C, C++, and Rust.
Fruit of an unprecedented collaboration between four major browser
vendors, \wasm ensures that even buggy or malicious code downloaded
from untrusted sources can be executed safely in a web browser~\cite{Haas:2017}.
To enforce security, \wasm programs are validated (type-checked) first and then
executed inside a sandbox that isolates untrusted code from the browser.
\emph{Memory safety} is key to the isolation mechanism of the
sandboxed execution environment: well-typed programs cannot corrupt
the memory outside the sandbox (e.g., the Javascript virtual machine).
Unfortunately, \wasm is still far from secure: buffer overflows and
use-after-free can still corrupt the memory of a program \emph{within}
the sandbox, opening the door to attacks like cross-site scripting and
remote code execution~\cite{Chasm:WASM}.
The presence of memory vulnerabilities in \wasm thwarts the strenuous efforts
devoted into securing unsafe languages like
C~\cite{checkedc,cets,ccured,ccured-toplas,Agten:2015} and developing
resource-aware memory-safe languages like Rust~\cite{Matsakis:2014,Jung:2017}.
Current compilers (e.g., Emscripten) do not attempt to protect compiled programs
from \wasm-level attackers exploiting well-known memory vulnerabilities.
Following the principled tradition of \emph{secure
compilation}~\cite{surv,Abadi:2012,rhc}, we propose to strengthen
the \wasm compilation chain with a provably secure
memory-safety-preserving compiler.
Fortunately, several aspects of \wasm promote rigorous reasoning and
help us in our study.
In particular, \wasm (1) has a (mostly) deterministic formal semantics that
rules out undefined behaviour and (2) is type-safe~\cite{Haas:2017}.
The specification of \wasm has even been mechanized and verified~\cite{Watt:2018}.
Furthermore, the existing security mechanisms of \wasm reduce the attack surface
available to target level attackers and thus simplify the job of our secure
compiler.
\wasm features \emph{structured control-flow} and separates code and data memory
segments, which, in combination, enforce coarse-grained control-flow
integrity~\cite{Abadi:2009, Abadi:2005} removing by construction classic
stack-smashing and return-oriented programming attacks.
In addition, \wasm provides state and memory \emph{encapsulation} through
modules, which represent natural boundaries where to enforce
security~\cite{Haas:2017}.
Assuming some degree of freedom when setting module boundaries, we believe
that a secure compiler could reuse the existing mechanisms of \wasm to
enforce memory safety at the target level, in the short term.
However, this approach rests on a strong assumption, namely that the
compiler has direct control over how code gets compartmentalized.
As this may not always be the case, and thus for a long-term solution,
we draw inspiration from Memory Safe WebAssembly (\mswasm), a recent
design proposal for extending \wasm with hardware-supported
progressive memory-safety capabilities~\cite{mswasm}.
A secure compiler relying on \mswasm language-level support for memory-safety enforcement could allow looser module
boundaries.

In the rest of this short paper, we discuss what notions of memory
safety we wish to enforce and how to formally express them as
(hyper)properties.\footnote{Properties are defined over single runs of
a program, while hyperproperties involve multiple
runs~\cite{ClarksonS10}.  } %
Then, we outline \mswasm and argue that it is a suitable target
candidate for secure compilation. %
Finally, we discuss which secure compilation criterion to
use when preserving memory safety to \mswasm. %

\section{Memory Safety as a (Hyper)Property}\label{sec:memsaf}
Establishing rigorous security guarantees for our compiler requires a
formal definition of memory safety, an intuitive notion that has been
surprisingly hard to pin down~\cite{ms-hicks-blog}.
The exact definition of memory safety has important ramifications for
our work because it determines what class of security properties our
compiler has to preserve and thus what protection mechanisms are
needed~\cite{rhc}.
Previous works on safe variants of
C~\cite{checkedc,cets,ccured,ccured-toplas,Agten:2015} treat memory
safety as a simple safety property enforceable by \emph{reference
monitors}~\cite{Schneider:2000} that detect specific memory violations
(e.g., accessing freed memory or an array out-of-bounds).
Seeking a definition that trascends bad behaviours, \citet{memsafety}
associate memory safety with reasoning principles about state akin to
non-inteferference~\cite{Goguen82}.
Since non-interference relates \emph{pairs} of executions, their
definition ascribes memory safety to the class of
2-hypersafety~\cite{ClarksonS10}, which is arguably harder to
preserve \emph{robustly} than safety~\cite{rhc,rsc}.
Here, we consider a notion of memory safety based on \emph{color
tags}, inspired by a line of work on micro-policies for tag-based
security monitors~\cite{Amorim15,Dhawan:2015}.
Briefly, memory locations and pointers are tagged with colors and
a memory violation occurs when a pointer accesses memory tagged with a
different color.
Unlike the definitions of the works mentioned above, this safety property
is trace-based and agnostic to
the specific semantics of the languages involved and their
syntax---the trace only contains memory relevant actions (i.e.,
memory allocation, free, read and write).
Furthermore, this definition let us study various relaxation of memory
safety that could describe precisely the progressive guarantees
of \mswasm, including spatial, (relaxed) temporal
safety\footnote{Relaxed temporal safety allows memory accesses through
dangling pointers as long as the memory pointed to has not been
reallocated~\cite{mswasm}.} and pointer integrity, as well as novel
properties that considers only data integrity.\footnote{To reduce the
overhead of enforcing memory-safety, some tools support modes that
check only memory writes \cite{softbound,Duck:2016,Duck:17}. }

\section{Memory Safe WebAssembly}\label{sec:pms}
Memory Safe WebAssembly (\mswasm) is an extension of \wasm designed to
capture sufficient metadata about pointers and memory regions to
enforce memory safety efficiently, levereging dedicated hardware.
In particular, \mswasm promotes a \emph{progressive enforcement} of
memory safety, i.e., depending on application-specific
security-performance trade-offs and what particular hardware is
available, the same abstractions can enforce ``weaker'' forms of
memory safety.
The core features of \mswasm design are \emph{segment memories}, i.e.,
linearly addressable, zero-initialized, manually managed extents of
memory, and \emph{handles}, i.e., possibly-corrupted (forged) pointers
enriched with bounds metadata.
To enforce memory safety, \mswasm restricts the interaction between
segments and memories appropriately (e.g., only handlers can access
segments, provided that they point within their bounds).
In order to use \mswasm as a target language in our secure compilation
chain, we have to first formalize its design and semantics.
Then, using variations of our trace-based definition of memory safety
from above, we intend to prove its progressive memory-safety
guarantees involving spatial, relaxed temporal safety, and pointer
integrity, and establish their relative strengths.
With the help of \mswasm abstractions, we are then going to design a
class of secure compilers that preserve clearly-defined notions of
memory safety.

\section{Secure Compilation to \mswasm}\label{sec:sc}
To establish the security guarantees of our compilers, we prove that
they attain a secure compilation criterion.
Then, to further clarify their security guarantees, we consider
general compilation criteria that preserve whole classes of security
properties (including memory-safety), instead of using an ad-hoc
criterion.
Given that we can express memory safety as a safety property as well
as a 2-hypersafety property, we adopt two of the robust compilation
criteria proposed by \citet{rhc}, namely \emph{Robust Safety Property
Preservation} (RSP) and \emph{Robust 2-Hypersafety Preservation}
(R2HSP).
Intuitively, these criteria require compilers to preserve
(hyper)properties of source programs even when they are compiled and
linked with arbitrary target code, thus protecting robustly against
all \emph{active} target-level attackers.
In practice, equivalent \emph{property-free} characterizations
simplify significantly the proofs of robust criteria
preservation~\cite{rhc}.
Specifically, for any compiled program and target context triggering a
bad behaviour, we have to find a corresponding source-level context
that produces the same bad behaviour.
To reconstruct suitable source-level contexts systematically, we can
apply known proof techniques based
on \emph{backtranslation}~\cite{rsc,rhc,NewBA16,DevriesePP16}.
The proofs of RSP and R2HSP differ mainly over the \emph{kind} of bad
behaviours involved, which are determined by the properties that they
preserve (safety and 2-hypersafety).
Since safety is a simple property, bad behaviours are just finite
 traces (prefixes) in RSP.
In R2HSP, bad behaviours consist of pair of prefixes because
 2-hypersafety is just a generalization of safety to a
 2-hyperproperty \cite{ClarksonS10}.
By including all memory-relevant actions in our traces, we gain
confidence that the criteria above characterizes correctly the class
of memory-safety-preserving compilers that we intend to study.

\newpage

{\small
\textbf{Acknowledgements:}
This work was partially supported by the German Federal Ministry of Education and Research (BMBF) through funding for the CISPA-Stanford Center for Cybersecurity (FKZ: 13N1S0762).

}

\bibliographystyle{plainnat}
\bibliography{biblio.bib}

\begin{thebibliography}{29}
\providecommand{\natexlab}[1]{#1}
\providecommand{\url}[1]{\texttt{#1}}
\expandafter\ifx\csname urlstyle\endcsname\relax
  \providecommand{\doi}[1]{doi: #1}\else
  \providecommand{\doi}{doi: \begingroup \urlstyle{rm}\Url}\fi

\bibitem[Abadi and Plotkin(2012)]{Abadi:2012}
Mart\'{\i}n Abadi and Gordon~D. Plotkin.
\newblock On protection by layout randomization.
\newblock \emph{ACM Trans. Inf. Syst. Secur.}, 15\penalty0 (2):\penalty0
  8:1--8:29, July 2012.
\newblock ISSN 1094-9224.
\newblock \doi{10.1145/2240276.2240279}.
\newblock URL \url{http://doi.acm.org/10.1145/2240276.2240279}.

\bibitem[Abadi et~al.(2005)Abadi, Budiu, Erlingsson, and Ligatti]{Abadi:2005}
Mart\'{\i}n Abadi, Mihai Budiu, \'{U}lfar Erlingsson, and Jay Ligatti.
\newblock A theory of secure control flow.
\newblock In \emph{Proceedings of the 7th International Conference on Formal
  Methods and Software Engineering}, ICFEM'05, pages 111--124, Berlin,
  Heidelberg, 2005. Springer-Verlag.
\newblock ISBN 3-540-29797-9, 978-3-540-29797-0.
\newblock \doi{10.1007/11576280_9}.
\newblock URL \url{http://dx.doi.org/10.1007/11576280_9}.

\bibitem[Abadi et~al.(2009)Abadi, Budiu, Erlingsson, and Ligatti]{Abadi:2009}
Mart\'{\i}n Abadi, Mihai Budiu, \'{U}lfar Erlingsson, and Jay Ligatti.
\newblock Control-flow integrity principles, implementations, and applications.
\newblock \emph{ACM Trans. Inf. Syst. Secur.}, 13\penalty0 (1):\penalty0
  4:1--4:40, November 2009.
\newblock ISSN 1094-9224.
\newblock \doi{10.1145/1609956.1609960}.
\newblock URL \url{http://doi.acm.org/10.1145/1609956.1609960}.

\bibitem[Abate et~al.(2019)Abate, Blanco, Garg, Hri\c{t}cu, Patrignani, and
  Thibault]{rhc}
Carmine Abate, Roberto Blanco, Deepak Garg, C\u{a}t\u{a}lin Hri\c{t}cu, Marco
  Patrignani, and J\'er\'emy Thibault.
\newblock Journey beyond full abstraction: Exploring robust property
  preservation for secure compilation.
\newblock In \emph{2019 IEEE 32th Computer Security Foundations Symposium}, CSF
  2019, June 2019.

\bibitem[Agten et~al.(2015)Agten, Jacobs, and Piessens]{Agten:2015}
Pieter Agten, Bart Jacobs, and Frank Piessens.
\newblock Sound modular verification of c code executing in an unverified
  context.
\newblock In \emph{Proceedings of the 42Nd Annual ACM SIGPLAN-SIGACT Symposium
  on Principles of Programming Languages}, POPL '15, pages 581--594, New York,
  NY, USA, 2015. ACM.
\newblock ISBN 978-1-4503-3300-9.
\newblock \doi{10.1145/2676726.2676972}.
\newblock URL \url{http://doi.acm.org/10.1145/2676726.2676972}.

\bibitem[Azevedo~de Amorim et~al.(2018)Azevedo~de Amorim, Hri{\c{T}}cu, and
  Pierce]{memsafety}
Arthur Azevedo~de Amorim, C{\u{a}}t{\u{a}}lin Hri{\c{T}}cu, and Benjamin~C.
  Pierce.
\newblock The meaning of memory safety.
\newblock In Lujo Bauer and Ralf K{\"u}sters, editors, \emph{Principles of
  Security and Trust}, pages 79--105, Cham, 2018. Springer International
  Publishing.
\newblock ISBN 978-3-319-89722-6.

\bibitem[Clarkson and Schneider(2010)]{ClarksonS10}
Michael~R. Clarkson and Fred~B. Schneider.
\newblock Hyperproperties.
\newblock \emph{Journal of Computer Security}, 18\penalty0 (6):\penalty0
  1157--1210, 2010.
\newblock \doi{10.3233/JCS-2009-0393}.
\newblock URL
  \url{https://www.cs.cornell.edu/~clarkson/papers/clarkson_hyperproperties_journal.pdf}.

\bibitem[d.~{Amorim} et~al.(2015)d.~{Amorim}, {Dénès}, {Giannarakis},
  {Hritcu}, {Pierce}, {Spector-Zabusky}, and {Tolmach}]{Amorim15}
A.~A. d.~{Amorim}, M.~{Dénès}, N.~{Giannarakis}, C.~{Hritcu}, B.~C. {Pierce},
  A.~{Spector-Zabusky}, and A.~{Tolmach}.
\newblock Micro-policies: Formally verified, tag-based security monitors.
\newblock In \emph{2015 IEEE Symposium on Security and Privacy}, pages
  813--830, May 2015.
\newblock \doi{10.1109/SP.2015.55}.

\bibitem[Devriese et~al.(2016)Devriese, Patrignani, and Piessens]{DevriesePP16}
Dominique Devriese, Marco Patrignani, and Frank Piessens.
\newblock Fully-abstract compilation by approximate back-translation.
\newblock In \emph{43nd Annual {ACM} {SIGPLAN-SIGACT} Symposium on Principles
  of Programming Languages}, 2016.

\bibitem[Dhawan et~al.(2015)Dhawan, Hritcu, Rubin, Vasilakis, Chiricescu,
  Smith, Knight, Pierce, and DeHon]{Dhawan:2015}
Udit Dhawan, Catalin Hritcu, Raphael Rubin, Nikos Vasilakis, Silviu Chiricescu,
  Jonathan~M. Smith, Thomas~F. Knight, Jr., Benjamin~C. Pierce, and Andre
  DeHon.
\newblock Architectural support for software-defined metadata processing.
\newblock In \emph{Proceedings of the Twentieth International Conference on
  Architectural Support for Programming Languages and Operating Systems},
  ASPLOS '15, pages 487--502, New York, NY, USA, 2015. ACM.
\newblock ISBN 978-1-4503-2835-7.
\newblock \doi{10.1145/2694344.2694383}.
\newblock URL \url{http://doi.acm.org/10.1145/2694344.2694383}.

\bibitem[Disselkoen et~al.(2019)Disselkoen, Renner, Watt, Garfinkel, Levy, and
  Stefan]{mswasm}
Craig Disselkoen, John Renner, Conrad Watt, Tal Garfinkel, Amit Levy, and Deian
  Stefan.
\newblock Position paper: Progressive memory safety for webassembly.
\newblock In \emph{Proceedings of the 8th International Workshop on Hardware
  and Architectural Support for Security and Privacy}, HASP '19, pages
  4:1--4:8, New York, NY, USA, 2019. ACM.
\newblock ISBN 978-1-4503-7226-8.
\newblock \doi{10.1145/3337167.3337171}.
\newblock URL \url{http://doi.acm.org/10.1145/3337167.3337171}.

\bibitem[Duck and Yap(2016)]{Duck:2016}
Gregory~J. Duck and Roland H.~C. Yap.
\newblock Heap bounds protection with low fat pointers.
\newblock In \emph{Proceedings of the 25th International Conference on Compiler
  Construction}, CC 2016, pages 132--142, New York, NY, USA, 2016. ACM.
\newblock ISBN 978-1-4503-4241-4.
\newblock \doi{10.1145/2892208.2892212}.
\newblock URL \url{http://doi.acm.org/10.1145/2892208.2892212}.

\bibitem[Duck et~al.(2017)Duck, Yap, and Cavallaro]{Duck:17}
Gregory~J. Duck, Roland H.~C. Yap, and Lorenzo Cavallaro.
\newblock Stack bounds protection with low fat pointers.
\newblock In \emph{24th Annual Network and Distributed System Security
  Symposium, {NDSS} 2017, San Diego, California, USA, February 26 - March 1,
  2017}, 2017.
\newblock URL
  \url{https://www.ndss-symposium.org/ndss2017/ndss-2017-programme/stack-object-protection-low-fat-pointers/}.

\bibitem[{Goguen} and {Meseguer}(1982)]{Goguen82}
J.~A. {Goguen} and J.~{Meseguer}.
\newblock Security policies and security models.
\newblock In \emph{1982 IEEE Symposium on Security and Privacy}, pages 11--11,
  April 1982.
\newblock \doi{10.1109/SP.1982.10014}.

\bibitem[Haas et~al.(2017)Haas, Rossberg, Schuff, Titzer, Holman, Gohman,
  Wagner, Zakai, and Bastien]{Haas:2017}
Andreas Haas, Andreas Rossberg, Derek~L. Schuff, Ben~L. Titzer, Michael Holman,
  Dan Gohman, Luke Wagner, Alon Zakai, and JF~Bastien.
\newblock Bringing the web up to speed with webassembly.
\newblock In \emph{Proceedings of the 38th ACM SIGPLAN Conference on
  Programming Language Design and Implementation}, PLDI 2017, pages 185--200,
  New York, NY, USA, 2017. ACM.
\newblock ISBN 978-1-4503-4988-8.
\newblock \doi{10.1145/3062341.3062363}.
\newblock URL \url{http://doi.acm.org/10.1145/3062341.3062363}.

\bibitem[Hicks(2014)]{ms-hicks-blog}
Michael Hicks.
\newblock What is memory safety?
\newblock \url{http://www.pl-enthusiast.net/2014/07/21/memory-safety/}, 2014.
\newblock Accessed: 2019-10-16.

\bibitem[Jung et~al.(2017)Jung, Jourdan, Krebbers, and Dreyer]{Jung:2017}
Ralf Jung, Jacques-Henri Jourdan, Robbert Krebbers, and Derek Dreyer.
\newblock Rustbelt: Securing the foundations of the rust programming language.
\newblock \emph{Proc. ACM Program. Lang.}, 2\penalty0 (POPL):\penalty0
  66:1--66:34, December 2017.
\newblock ISSN 2475-1421.
\newblock \doi{10.1145/3158154}.
\newblock URL \url{http://doi.acm.org/10.1145/3158154}.

\bibitem[Matsakis and Klock(2014)]{Matsakis:2014}
Nicholas~D. Matsakis and Felix~S. Klock, II.
\newblock The rust language.
\newblock In \emph{Proceedings of the 2014 ACM SIGAda Annual Conference on High
  Integrity Language Technology}, HILT '14, pages 103--104, New York, NY, USA,
  2014. ACM.
\newblock ISBN 978-1-4503-3217-0.
\newblock \doi{10.1145/2663171.2663188}.
\newblock URL \url{http://doi.acm.org/10.1145/2663171.2663188}.

\bibitem[Nagarakatte et~al.(2009)Nagarakatte, Zhao, Martin, and
  Zdancewic]{softbound}
Santosh Nagarakatte, Jianzhou Zhao, Milo~M.K. Martin, and Steve Zdancewic.
\newblock Softbound: Highly compatible and complete spatial memory safety for
  c.
\newblock \emph{SIGPLAN Not.}, 44\penalty0 (6):\penalty0 245--258, June 2009.
\newblock ISSN 0362-1340.
\newblock \doi{10.1145/1543135.1542504}.
\newblock URL \url{http://doi.acm.org/10.1145/1543135.1542504}.

\bibitem[Nagarakatte et~al.(2010)Nagarakatte, Zhao, Martin, and
  Zdancewic]{cets}
Santosh Nagarakatte, Jianzhou Zhao, Milo~M.K. Martin, and Steve Zdancewic.
\newblock Cets: Compiler enforced temporal safety for c.
\newblock \emph{SIGPLAN Not.}, 45\penalty0 (8):\penalty0 31--40, June 2010.
\newblock ISSN 0362-1340.
\newblock \doi{10.1145/1837855.1806657}.
\newblock URL \url{http://doi.acm.org/10.1145/1837855.1806657}.

\bibitem[{NCC Group Whitepaper}(2018)]{Chasm:WASM}
{NCC Group Whitepaper}.
\newblock
  \url{https://i.blackhat.com/us-18/Thu-August-9/us-18-Lukasiewicz-WebAssembly-A-New-World-of-Native_Exploits-On-The-Web-wp.pdf},
  2018.
\newblock Accessed: 2019-10-11.

\bibitem[Necula et~al.(2005{\natexlab{a}})Necula, Condit, Harren, McPeak, and
  Weimer]{ccured}
George~C. Necula, Jeremy Condit, Matthew Harren, Scott McPeak, and Westley
  Weimer.
\newblock Ccured: Type-safe retrofitting of legacy software.
\newblock \emph{ACM Trans. Program. Lang. Syst.}, 27\penalty0 (3):\penalty0
  477--526, May 2005{\natexlab{a}}.
\newblock ISSN 0164-0925.
\newblock \doi{10.1145/1065887.1065892}.
\newblock URL \url{http://doi.acm.org/10.1145/1065887.1065892}.

\bibitem[Necula et~al.(2005{\natexlab{b}})Necula, Condit, Harren, McPeak, and
  Weimer]{ccured-toplas}
George~C. Necula, Jeremy Condit, Matthew Harren, Scott McPeak, and Westley
  Weimer.
\newblock Ccured: Type-safe retrofitting of legacy software.
\newblock \emph{ACM Trans. Program. Lang. Syst.}, 27\penalty0 (3):\penalty0
  477--526, May 2005{\natexlab{b}}.
\newblock ISSN 0164-0925.
\newblock \doi{10.1145/1065887.1065892}.
\newblock URL \url{http://doi.acm.org/10.1145/1065887.1065892}.

\bibitem[New et~al.(2016)New, Bowman, and Ahmed]{NewBA16}
Max~S. New, William~J. Bowman, and Amal Ahmed.
\newblock Fully abstract compilation via universal embedding.
\newblock In \emph{21st {ACM} {SIGPLAN} International Conference on Functional
  Programming, {ICFP}}, pages 103--116, 2016.
\newblock \doi{10.1145/2951913.2951941}.
\newblock URL \url{https://www.williamjbowman.com/resources/fabcc-paper.pdf}.

\bibitem[Patrignani and Garg(2019)]{rsc}
Marco Patrignani and Deepak Garg.
\newblock {Robustly Safe Compilation}.
\newblock In \emph{Programming Languages and Systems - 28th European Symposium
  on Programming, {ESOP} 2019}, ESOP'19, 2019.

\bibitem[Patrignani et~al.(2019)Patrignani, Ahmed, and Clarke]{surv}
Marco Patrignani, Amal Ahmed, and Dave Clarke.
\newblock Formal approaches to secure compilation a survey of fully abstract
  compilation and related work.
\newblock \emph{ACM Comput. Surv.}, 51\penalty0 (6):\penalty0 125:1--125:36,
  January 2019.
\newblock \doi{10.1145/3280984}.
\newblock URL \url{https://doi.org/10.1145/3280984}.

\bibitem[Ruef et~al.(2019)Ruef, Lampropoulos, Sweet, Tarditi, and
  Hicks]{checkedc}
Andrew Ruef, Leonidas Lampropoulos, Ian Sweet, David Tarditi, and Michael
  Hicks.
\newblock Achieving safety incrementally with checked c.
\newblock In \emph{Principles of Security and Trust}, pages 76--98, Cham, 2019.
  Springer International Publishing.
\newblock ISBN 978-3-030-17138-4.

\bibitem[Schneider(2000)]{Schneider:2000}
Fred~B. Schneider.
\newblock Enforceable security policies.
\newblock \emph{ACM Trans. Inf. Syst. Secur.}, 3\penalty0 (1):\penalty0 30--50,
  February 2000.
\newblock ISSN 1094-9224.
\newblock \doi{10.1145/353323.353382}.
\newblock URL \url{http://doi.acm.org/10.1145/353323.353382}.

\bibitem[Watt(2018)]{Watt:2018}
Conrad Watt.
\newblock Mechanising and verifying the webassembly specification.
\newblock In \emph{Proceedings of the 7th ACM SIGPLAN International Conference
  on Certified Programs and Proofs}, CPP 2018, pages 53--65, New York, NY, USA,
  2018. ACM.
\newblock ISBN 978-1-4503-5586-5.
\newblock \doi{10.1145/3167082}.
\newblock URL \url{http://doi.acm.org/10.1145/3167082}.

\end{thebibliography}
\end{document}